\documentclass[11pt]{article}
\usepackage{times}
\usepackage{graphics}
\usepackage{amsmath}
\usepackage{amssymb}
\usepackage{ifthen}
\usepackage{calc}
\usepackage{hyperref}
\usepackage{color}

\newcommand{\nlogn}[1]{\ensuremath{\log \ifthenelse{\equal{#1}{1}}{}{^{#1}}n}}

\newcommand{\myenumlist}{
  \begin{list}{\arabic{enumi}.}
              {\usecounter{enumi}
               \setlength{\labelwidth}{\myenumlabwid}
               \setlength{\labelsep}{\mylabelsep}
               \setlength{\leftmargin}{\myenumlabwid}
               \addtolength{\leftmargin}{\mylabelsep}
               \setlength{\topsep}{\mytopsep} 
               \setlength{\parskip}{0in}
               \setlength{\itemsep}{\myitemsep} 
               \setlength{\partopsep}{0in} }     }

\newcommand{\vio}[1]{\ensuremath{\widehat{#1}}}
\newcommand{\flow}[1]{\ensuremath{\widehat{#1}_f}}
\newcommand{\ellp}[1]{\ensuremath{L_{0,#1}}}
\newcommand{\wellp}[1]{\ensuremath{L_{0,#1|C}}}
\newcommand{\trimerr}[1]{\ensuremath{\mathrm{t\_err}(#1)}}
\newcommand{\terr}{\ensuremath{\mathrm{t\_err}}}
\newcommand{\winlow}[1]{\ensuremath{w_{C \prec}(#1)}}
\newcommand{\winhigh}[1]{\ensuremath{w_{C \succ}(#1)}}

\newtheorem{theorem}{Theorem}
\newtheorem{corollary}[theorem]{Corollary}
\newtheorem{proposition}[theorem]{Proposition}
\newtheorem{lemma}[theorem]{Lemma}

\begin{document}

\begin{center}
{\Large \textbf{$L_0$ Isotonic Regression With Secondary Objectives}}
\bigskip

{\large Quentin F. Stout}\\[\medskipamount]
qstout@umich.edu\\
www.eecs.umich.edu/{\,\raisebox{-0.5ex}{\textasciitilde}qstout/}

\end{center}

 \bigskip

\noindent \textbf{Abstract:~}
 We provide algorithms for isotonic regression minimizing  $L_0$ error (Hamming distance).
This is also known as monotonic relabeling, and is applicable when labels have a linear ordering but not necessarily a metric.
There may be exponentially many optimal relabelings, so we look at secondary criteria to determine which are best.
For arbitrary ordinal labels the criterion is maximizing the number of labels which are only changed to an adjacent label (and recursively apply this).
For real-valued labels we minimize the $L_p$ error.
For linearly ordered sets we also give algorithms which minimize the sum of the $L_p$ and weighted $L_0$ errors,  a form of penalized (regularized) regression.
We also examine $L_0$ isotonic regression on multidimensional coordinate-wise orderings.
\medskip 

\bigskip

\section{Introduction} \label{sec:intro}

There are many scenarios when one expects that an attribute of objects increases as a function of other attributes.
For example, if people's weight is categorized as \{underweight, normal, overweight, obese\}, one would expect that if the daily intake of fat is held constant then the categorization increases as the carbohydrate intake increases, and similarly it increases if the daily carbohydrate intake is held constant and the fat intake increases.
No assumptions are made about the weight of a person with higher fat, but lower carbohydrate, vs.\ one with lower fat but higher carbohydrates.
However, noisy datasets may have data that violates these assumptions.
Constructing a representation of the data which obeys the assumptions and has minimal changes to the data is generically known as isotonic regression.
In this example the dependent variable, weight categorization, is a label with no assumed metric, only an ordering, and the problem is more commonly known as monotonic relabeling.

As datasets become increasingly complex, often with only ordinal ordering on some attributes, such nonparametric regressions become increasingly more useful, but are often nonunique and difficult to compute.
We will use optimizations based on secondary criteria to choose among the possible regressions, and give algorithms to find the resulting regressions.

More precisely, let $V$ be a set with a partial order $\prec$, and $\mathcal{L}$ be a linearly ordered set.
A \textit{label function} $f$ is a mapping $f: V \to  \mathcal{L}$. 
Given label functions $f, g$, the $L_0$ distance between them, $\|f-g\|_0$, is 
$$
\sum_{v \in V} \mathbf{1}\cdot(f(v) \neq g(v))
$$
This is also known as the \textit{Hamming distance}, 0-1 loss, or Kronecker delta loss.
It is similar to the well-known $L_p$ distance when the labels are real values:
$$
\|f-g\|_p =
\left\{
\begin{array}{ll} 
  \left(\sum_{v \in V} |f(v) -g(v)|^p\right)^{1/p} & 1 \leq p < \infty \smallskip \\
  \max_{v \in V} |f(v) - g(v)|                     & p = \infty
\end{array}
\right.
$$
Here we consider two classes of label values: one is when $\mathcal{L} = \{\lambda(1), \ldots \lambda(\ell)\}$ for a positive integer $\ell$, where $\lambda(1) < \ldots < \lambda(\ell)$.
The other is where $\mathcal{L}$ is the real numbers.
The former is used in papers which describe the problem as ``relabeling", such as~\cite{Relab_Radetal12}, while the latter is used in papers on function approximation and penalized regression.
Our results are applicable in both settings.

A label function $g$ is \textit{monotonic} (\textit{isotonic}) if whenever $u \prec v$, for $u,v \in V$, then $g(u) \leq g(v)$, i.e., it is a weakly order-preserving mapping from $V$ to $\mathcal{L}$.
Let $\Delta_p (f) = \min\{||f-g||_p: g$ is monotonic\}.
This is the \textit{$L_p$ distance of $f$ to monotonicity}.
The term \textit{distance to monotonicity}, without mention of $p$, typically means $L_0$ distance, and often the goal is to quickly estimate it, rather than determine it exactly~\cite{IsotonicTesting,LIS_SublinApproxMultidim,LIS_SublinApproxPosets,LIS_SublinApproxDyn1D_21v2}.
Such estimation, or the mere decision if a function is monotonic, has often been used in property testing~\cite{BelovsPropertyTest,IsotonicTesting,PropertyTest20,MonotonicityTesting13,BooleanFunctionTesting,LIS_SublinApproxMultidim,LocalMonotone10}.
Here we are interested in the exact value of $\Delta_0(f)$.

A monotonic function $g$ is an \textit{$L_p$-optimal monotonic relabeling}, or an \textit{$L_p$ isotonic regression}, of label function $f$ iff $||f-g||_p = \Delta_p(f)$.
For $p \geq 1$ this requires that the labels are real values.
$L_0$-optimal monotonic relabelings need not be unique, e.g., on a sequence of just 2 points, if $f$ is  \texttt{\small overweight, normal}, then~ \texttt{\small normal, normal} ~and~ \texttt{\small overweight, overweight} ~are both optimal monotonic relabelings.
Since optimal $L_0$ monotonic relabelings are not always unique, how should one decide which to use?
Here we decide by using $L_1$, $L_2$, or $L_\infty$ distances to select among them.
This question was also raised in~\cite{Relab_Radeeta09}, where a similar selection approach was used, but theirs required the regression values to be in a fixed finite set $\mathcal{L}$, while we consider arbitrary real-valued regressions as well.
We also give more efficient algorithms for the cases they considered.

When $f$ is real-valued and $p \in \{1, 2, \infty\}$ three scenarios are considered:
\begin{enumerate}
\item Find an isotonic function $g$ such that $g \in \arg\min \{||f-h||_p: h \mathrm{~isotonic~and~} ||f-h||_0=\Delta_0(f)\}$. Note that $g$ is an $L_0$ optimal isotonic relabeling of $f$.
$g$ will be called \textit{\ellp{p} optimal}.

\item Given a subset $C$ of vertices on which $f$ is isotonic, where $|C|  = n - \Delta_0(f)$, find an isotonic function $g$ such that $g \in \arg\min\,\{||f-h||_p: h \mathrm{~isotonic~and~} h=f \mathrm{~on~} C\}$.
Note that for any function $g^\prime$ optimizing  the criterion in 1) there is a $C^\prime$ of size $n-\Delta_0(f)$ where $f$ is isotonic on $C^\prime$ and $g^\prime$ optimizes the criterion here for $C^\prime$.
However, there may be $C$ for which a function optimizing the criterion here does not optimize that in 1).
$g$ will be called \textit{$\wellp{p}$ optimal}, or \textit{weakly $\ellp{p}$ optimal}.

\item Given $\alpha > 0$, find an isotonic function $g$ such that $g \in \arg\min \{||f-h||_p+\alpha ||f-h||_0: h \mathrm{~isotonic}\}$.
Note that it may be that $||f-g||_0 > \Delta_0(f)$, while if $\alpha$ is sufficiently large $g$ is in \ellp{p}.
\end{enumerate}
When the labels $\mathcal{L}$ are a finite set of real numbers we consider variants of 1) and 2) where all isotonic regressions must be $\mathcal{L}$-valued.
We also introduce strong and weak versions of $\mathcal{L}$-valued functions with  which iteratively optimize $L_0$ for a decreasing set of vertices.
In this case $\mathcal{L}$ merely needs to be ordered, not a subset of the reals.

We give algorithms for all of these when the dag is a linear order, but only for some of them when the dag is more complex.

As will be shown in Section~\ref{sec:background}, for an arbitrary dag most of the computational work involves a ``violator dag'', a directed acyclic graph that represents where the monotonicity condition is violated.
To help keep track of which ordering or dag is being used, we use $G=(V,E)$ to denote the initial dag and $\vio{G}=(\vio{V},\vio{E})$ to denote the violator dag.
We assume $V$ is connected, and if it isn't then the construction is used on each component independently.
Let $n= |V|$, $m=|E|$, $\vio{n}=|\vio{V}|$ and $\vio{m}=|\vio{E}|$.

By \textit{multidimensional ordering} or \textit{$d$-dimensional ordering} we mean that the vertices are from some $d$-dimensional space, where each dimension has a linear ordering, and the vertices are ordered by component-wise ordering,
i.e., for $x = (x_1,\ldots,x_d)$ and $y= (y_1,\ldots,y_d)$, $x \prec y$ iff $x \neq y$ and $x_i \leq y_i$ for all $1 \leq i \leq d$.
This is also known as the product order of linear orders.
While in general violator graphs may have $\vio{m} = \Theta(n^2)$,
in Section~\ref{sec:multidim} we use a Steiner dag (a dag with additional vertices) as the violator graph for $d$-dimensional points, with $\vio{m} = \tilde{\Theta}(n)$.
The size of the violator graph, and time required to construct it, are significantly smaller than previous algorithms, and these orderings seem to be the ones of most interest in many applications.

In many applications, for a label set $\mathcal{L}$ of size $\ell$, $\ell$ might be quite small and independent of $n$, such as in the \{underweight, normal, overweight, obese\} example, and so some of the results are stated in terms of $\ell$.
To simplify exposition we assume that the labels are $1 \ldots \ell$, but for $L_p$ secondary optimization where the labels are real numbers we use the true values.

We do not examine any specific application, concentrating instead on the algorithms and reducing the time so that they can be applied to large problems.
Previous algorithms~\cite{Relab_Feelders06,Relab_LinLinearHam,Relab_Radetal12} are reviewed in Section~\ref{sec:background}.
Examples and applications of $L_0$ isotonic regressions and monotonic relabeling appear in many papers, such as~\cite{DuivFeeldersNNclass,Relab_Feelders06,Relab_PijlsPot14,Relab_Radeeta09,Relab_Radetal12,RULEM17}.
There are many applications of it to monotonic learning and classification, see~\cite{Brabantetal,LearnRelab_Cano18} and the extensive references therein.
However, in some classification settings one must be careful about how monotonicity is applied~\cite{MonoDataSci_DeBaets19}.

\section{Background} \label{sec:background}

There are some special cases which have simple solutions.
Minimizing the number of changed labels is equivalent to maximizing the number of unchanged ones, and hence if $\prec$ is a simple linear ordering then the problem is the same as finding a nondecreasing subsequence of maximum length.
This well-known problem has an easy $\Theta(n \log \ell)$ solution.
When $\prec$ is an arbitrary order but $\ell = 2$ then the problem is equivalent to $\{0,1\}$-valued $L_1$ isotonic regression, for which numerous algorithms are known~\cite{AhujaOrlin,L1Iso_AHKW,Flow_BLLSSSW,Flow_CKLPPS,QPartition,QMultidim}.
These algorithms vary widely in their time and construction as a function of the underlying order.
If it is linear or a 2-dimensional grid then the regression can be found in $\Theta(n)$ time, an arbitrary set of vertices in $d$ dimensions with $d$-dimensional ordering in $\Theta(n^{1.5} \log^{d-1} n)$ time, and an arbitrary dag in $\Theta( nm + n^2 \log n)$ time.
The first three of these results appear in~\cite{QPartition,QMultidim}, while that for arbitrary dags appears in~\cite{L1Iso_AHKW}.
For integer-valued weights recent results have lowered the time for arbitrary dags to nearly linear in $m$~\cite{Flow_BLLSSSW,Flow_CKLPPS}, where the time is achieved with high probability. 

Similarly there is a range of times and algorithms for $L_0$ isotonic regression depending on the underlying dag, but an extra aspect, namely secondary criteria, causes yet more complexity.
The general case, without such criteria, was analyzed in~\cite{Relab_Feelders06,Relab_PijlsDilworth13,Relab_Radeeta09}, while here we introduce a faster algorithm for multidimensional vertices (Section~\ref{sec:multidim}) and various algorithms optimizing secondary criteria.

Many of the algorithms utilize maximum flow algorithms.
The time of our algorithms is analyzed in terms of  $\mathcal{T}(\check{n},\check{m})$, the time of the flow algorithm on a graph of $\check{n}$ vertices and $\check{m}$ edges.
We use $\tilde{\Theta}( \cdot )$ to denote that logarithmic factors in terms of $\check{n}$ are omitted.
Algorithms in~\cite{Relab_Feelders06,Relab_PijlsDilworth13,Relab_Radeeta09}  had $\mathcal{T}=\Theta(n^3)$.
This is what occurs when you apply the faster of Orlin's algorithm and King-Rao-Tarjan's to the standard violator dag since there the worst case has $\check{n} = \Theta(n)$ and $\check{m} = \Theta(n^2)$.
Faster flow algorithms are now known, though with some constraints on the flow capacities and/or probability of attaining the expected time.
If all weights and capacities are integers in the range $[1,U]$ then
the BLLSSSW algorithm~\cite{Flow_BLLSSSW} takes $\tilde{O}\big(\check{m}+\check{n}^\frac{3}{2} \log U \big)$  time
and the  CKLPPS algorithm~\cite{Flow_CKLPPS} takes $O(\check{m}^{1+o(1)} \log^2 U)$ time, both with high probability.
From now on we make the common assumption that $U$ at most polynomial in $n$, and hence all of the terms involving $U$ are absorbed in the $\tilde{O}$ notation, and thus currently $\mathcal{T}$ is no worse than $\tilde{O}(\min\{\check{m}^{1+o(1)}, \check{n}^\frac{3}{2}\})$ with high probability.

Flow algorithms are still undergoing improvement, and the wikipedia page~\cite{Flow_wiki} is usually up-to-date with the latest improvements.
In practice simpler, but asymptotically slower, algorithms are faster on small datasets, but we are more interested in performance for the ever increasing size of data collections.
Since the time of our algorithms is expressed in terms of $\mathcal{T}$ one can interpret them in terms of whatever flow algorithm is used, not necessarily the asymptotically fastest.

\begin{figure} 
\begin{enumerate}

\item Create a violator dag $\vio{G}=(\vio{V},\vio{E})$, where for all $u, v \in V$, there is a path from $u$ to $v$ in $\vio{G}$ iff $u \prec v$ and $f(u) > f(v)$.~~\{Note: $V \subset \vio{V}$.\}
\item Find a maximum antichain $C$ of $\vio{G}$
  \begin{enumerate}
    \item Create a flow graph $\flow{G}$ from $\vio{G}$.
    \item Find a minimum flow on $\flow{G}$ and use this to determine $C$.
  \end{enumerate}
\item Determine $f^\prime$:  
\begin{enumerate}
  \item For all $v \in V$ determine the window $[\winlow{v}, \winhigh{v}]$ induced by $C$. 
  \item Let $f^\prime$ be any isotonic function on $V$ that goes through the windows.
     Since $\winlow{v}= \winhigh{v} = f(v)$ for $v \in C$, $f^\prime = f$ on C.
\end{enumerate} 

\end{enumerate}

\caption{$L_0$-optimal monotonic relabeling $f^\prime$ of label function $f$ on $G=(V,E)$ with order $\prec$  (see \cite{Relab_Feelders06,Relab_PijlsDilworth13,Relab_Radetal12})  }
\label{fig:general}
\hrulefill
\end{figure}

For the general case of $L_0$ isotonic regression on arbitrary dags, \cite{Relab_Feelders06,Relab_PijlsDilworth13,Relab_Radeeta09} used an approach based on violating pairs.
Given a label function $f$, vertices $p, q \in V$ are a \textit{violating pair} iff $p \prec q$ and $f(p) > f(q)$, i.e., they violate the monotonicity requirement.
Let $p \prec_v q$ denote that $p, q$ are a violating pair.
Note that $\prec_v$ is transitive, and hence is a partial ordering on the vertices in $V$.
Their approach is outlined in Figure~\ref{fig:general}.
Detailed proofs appear in their papers, but the basic idea is to create a dag based on violating pairs and find an antichain of maximum size in it.
Let $\vio{G}=(V,\vio{E})$ be a dag where there is a path from $p$ to $q$ in $\vio{G}$ iff $p \prec_v q$ in $G$.
$\vio{G}$ is a \textit{violator dag}.
While $V$ is well-defined, there may be multiple sets of edges which give the same partial ordering.
The construction of $\vio{G}$ in step 1) will be analyzed later.

Vertices in $V \setminus \vio{V}$ are not in any violating pair, i.e., on them $f$ is monotonic with respect to all other vertices, and hence in any optimal relabeling their label does not change.
Step 2) determines those elements $C$ of $\vio{V}$ where the label will not be changed.
\cite{Relab_Feelders06,Relab_Radetal12} borrow an idea from \cite{Mohring} to determine $C$.
No two elements of $C$ can be a violating pair, i.e., are not comparable in the $\prec_v$ ordering.
A set of incomparable elements in a partial order is an antichain,
and hence in an optimal relabeling $C$ is a maximum antichain.

Here, rather than emphasizing the antichain's graph properties we look at its subsequent role in constructing an  isotonic function.
A set $C \subseteq V$ is \textit{$f$-isotonic} iff $f$ is isotonic on $C$.
Thus  a set of vertices is an antichain of maximum size iff it is an $f$-isotonic set of maximum size on the set of vertices that are in violator pairs.
From now on we will slightly abuse terminology and say that $C$ is an $f$-isotonic set of maximum size if it is an $f$-isotonic set of maximum size on the set of vertices that are in violator pairs, where $C$ also implicitly contains all vertices not in any violator pair.
An isotonic function $g$ is \textit{consistent} with an $f$-isotonic set $C$ iff $g=f$ on $C$.

\begin{figure}

\centerline{\scalebox{0.85}{\includegraphics{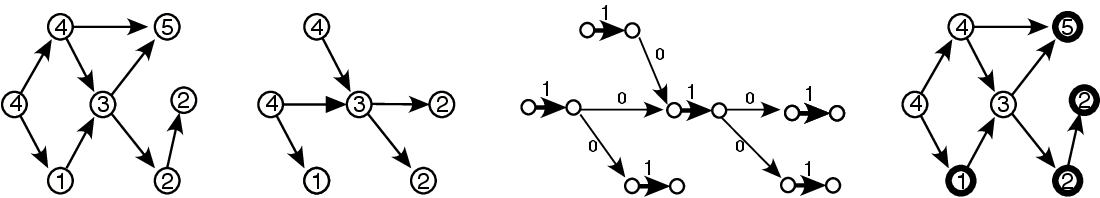}}}

\caption{Label function on dag $G$,~ one of its violator dags $\vio{G}$,~ flow graph $\flow{G}$,~ resulting $C$}
\label{fig:flowgraph}

\hrulefill
\end{figure}

Mohring~\cite{Mohring} showed how to find $C$ via minimal flows.
Figure~\ref{fig:flowgraph} shows a dag $G$ and one of its violator dags $\vio{G}$, which happens to be its transitive reduction.
$\vio{G}$ is transformed into a flow graph $\flow{G}$ as follows: each vertex $u$ of $\vio{G}$ is replaced by two vertices $u^{\mathrm{in}}, u^{\mathrm{out}}$, where each incoming edge to $u$ becomes an incoming edge to $u^{\mathrm{in}}$ with
0 minimum flow, each outgoing edge from $u$ becomes an outgoing edge from $u^{\mathrm{out}}$ with minimum flow 0, and there is an edge from $u^{\mathrm{in}}$ to $u^{\mathrm{out}}$ with minimum flow $1$.

Let $F$ be a minimum flow on $\flow{G}$, i.e., a flow where on each edge is at least the weight on that edge, and the total flow is as small as possible.
From $F$ one can determine a maximal cut, and the edges of weight 1 in this cut correspond to $C$.
The size of $C$ is the flow, and it is straightforward to show that $C$ is an antichain.
Thus the time of step 3) is the time to find a minimal flow on $\flow{G}$.
The flow is at most $|\vio{V}|$ and 
$\flow{G}$ has $\Theta(|\vio{G}|)$ vertices and $\Theta(|\vio{E}|)$ edges.

Step 3) determines $f^\prime$.
At each such vertex $v$ there is a ``window'' of label values $[b(v),t(v)]$ into which the new label must fall.
This is the range of values that are monotonically compatible with the values in $V^\prime$, i.e., the values at vertices where the function will not change.
The windows are isotonic in that the bottom values are an isotonic function on $V$, as are the top values.
Any isotonic function within these windows (such as always using the lower bound) can be used as $f^\prime$, and simple topological sort can be used to finish this step in $\Theta(\vio{m})$ time.
However, one may want to find an $f^\prime$ optimizing secondary objectives, the main contribution of this paper, which adds complexity.

Returning to step 2), there are several ways $\vio{G}$ can be created.
If $\prec$ is given explicitly via dag $G=(V,E)$ then $\vio{G}$ can be constructed via topological sorting, taking $\Theta(nm)$ time, or via matrix multiplication, taking $\Theta(n^\omega)$ time, where $\omega$ is such that matrix multiplication over the integers can be done in $\Theta(n^\omega)$ time.
For multidimensional vertices $\vio{G}$ can be constructed by pairwise comparisons, taking $\Theta(n^2)$ time, and in Section~\ref{sec:multidim} it is shown that this time can be reduced.

While a dag specifies a partial ordering, there may be multiple dags specifying the same partial ordering.
For example, a dag with an edge for each pair of vertices in the transitive closure, and one with an edge for each pair in the transitive reduction, specify the same partial order, but the latter may be significantly smaller.
Both can be constructed in $\Theta(\min\{\vio{n} m, \vio{n}^\omega\})$ time, and 
therefore \cite{Relab_Radetal12} uses the transitive reduction for $\vio{G}$.

If the data is almost in order and there are few violating pairs then $\vio{G}$ may be much smaller than $G$.
However, in the worst case even the transitive reduction of $\vio{G}$ can be quite large.
For example, on the linear order on $V=\{1, 2, \ldots, n\}$, for $n$ even, suppose $f$ is 
$\frac{n}{2}\!+\!1, \frac{n}{2}\!+\!2,  \ldots n, 1, 2 \ldots \frac{n}{2}$.
Then for $p,q \in V$, $p \prec_v q$ iff $p\in[1,\frac{n}{2}]$ and $q \in [1\!+\!\frac{n}{2},\,n]$.
The transitive closure and reduction are the same, with $\vio{n}=n$ and $\vio{m}=n^2/4$.
Here $\vio{m} \gg m$.
Further, there can be exponentially maximum $f$-isotonic sets, as 2, 1, 4, 3, 6, 5, \ldots shows since one needs to choose 1 from each pair of the form $k$, $k-1$.

Note that \cite{L1Iso_AHKW} gives an $L_1$ isotonic regression algorithm that is based on violating pairs and flows but does not require constructing a potentially large violator graph.

\section{Secondary Optimality}  \label{sec:secondary}

For general label functions there can be many maximum $f$-isotonic sets, and for each there may be many possible values for the remaining vertices as long as they satisfy monotonicity.
For example, given label values \texttt{\small small $<$ medium $<$ large $<$ X-large}, for a linearly ordered set with data values \texttt{\small large, medium, small}, any one of these values provides a maximum $f$-isotonic set.
However, many researchers might prefer using \{\texttt{\small medium}\}, with regression values \texttt{\small medium, medium, medium}.
However, even with this choice of maximum $f$-isotonic set, \texttt{\small small, medium, large} is also $L_0$ optimal, as is \texttt{\small small, medium, X-large}.
This raises the questions: which maximum $f$-isotonic set should be used, and what should the regression values be for the remaining vertices?
This question was also raised in~\cite{Relab_Radeeta09}.

When the labels are real numbers one way to choose the best $L_0$ regression is to minimize $L_p$ error.
That is, given $1 \leq p \leq \infty$, for a given label function $f$ choose an isotonic function 
$g \in \arg\min\{||f-h||_p: h~\mathrm{isotonic}, ||f-h||_0 = \Delta_0(f)\}$.
We call this \textit{strong optimality}, denoted $\ellp{p}(f)$.
A somewhat weaker optimality is to first choose an $f$-isotonic set $C$ of maximum size and then minimize the $L_p$ error of all isotonic functions which agree with $f$ on $C$, i.e., find a $g \in \arg\min\{||f-h||_p: h \mathrm{~isotonic}, f=h~\mathrm{~on~} C\}$.
We call this \textit{weak optimality}, denoted $\wellp{p}(f)$.
Note that for any strongly optimal $g \in \ellp{p}(f)$ there is an $f$-isotonic set $C$ of maximal size such that $g \in \wellp{p}(f)$.
For example, for function values 3, 2, 1 on a linear order, for $1 \leq p \leq \infty$, using $C = \{x\}$, $x \in \{1, 2, 3\}$, the isotonic regression function $x$, $x$, $x$ is in $\wellp{p}(f)$, and 2, 2, 2 is in $\ellp{p}(f)$.

In general finding weakly optimal functions is easier than finding strongly optimal ones.
Throughout we examine the values of $p$ most widely utilized, namely $p \in \{1, 2, \infty\}$.
More generally one could consider $L_{p,q}$, optimizing the $L_q$ norm among all isotonic regressions that optimize the $L_p$ norm.
For $1 < p < \infty$ this is not interesting since there is a unique $L_p$ isotonic regression.
For $p=1$ there may be some variation possible, and  for $p=\infty$ there are often many optimal regressions, much like $L_0$.
However, this general setting will not be considered here.

One may want $L_0$ as a secondary criterion, not just the primary one, an approach that can be used when the labels merely have an ordering with no metric.
For a weak version with a fixed set $\mathcal{L}$ of labels let $C_0$ be an $f$-isotonic set of maximal size.
Regression values at vertices not in $C_0$ must be 1 or more labels away from their original value.
Among these vertices find an $f$-isotonic set $C_1$ of maximum size where all regression values change to adjacent values, and for these vertices pin their values to their trimmed value.
This shrinks the windows, and now all vertices that aren't pinned must have regression values 2 or more away from their trimmed value.
Find an $f$-isotonic set $C_2$ fitting within the windows which maximizes the number of vertices 2 labels away from their trimmed value, etc.
An isotonic regression obtained this way is called a  \textit{weak} \wellp{0} regression.
While the size of $C_0$ is unique, the size of $C_1$ depends on the choice of $C_0$, the size of $C_2$ depends on $C_0$ and $C_1$, etc.

When the labels are real-valued \wellp{0} is not well defined.
Instead, for arbitrary real-valued labels let $g$ be an isotonic function where all its values are label values, and let $(e_1, e_2, \ldots e_n)$ be the values $\{ |f(v)-g(v)| : v \in V\}$ in sorted order.
Viewing this as a sequence of $n$ values, a \textit{strong} \ellp{0} isotonic regression is one which has a sequence lexically first among all isotonic functions (this may not be unique).
We call an $L_0$ isotonic regression strong if it is a strong \ellp{0} function.

\subsection{Label- vs.\ Real- Valued Regressions}

For $\ellp{p}$ or $\wellp{p}$ isotonic regression, $1 \leq p$, in some settings the values of the initial function $f$ must be from a set $\mathcal{L}$ of real-valued labels, while in others they can be arbitrary real numbers.
Further, in the former one may want the values of the isotonic regression to be from $\mathcal{L}$ or from arbitrary reals.
Since most isotonic regression algorithms are developed for regression values being arbitrary real numbers, or integers, when the values must be from $\mathcal{L}$ the values of an optimal arbitrary real-valued regression need to be converted to labels.
Let $g$ be an optimal real-valued isotonic regression for whatever metric is being considered.
To construct a $g^\prime$ which is an optimal $\mathcal{L}$-valued regression at vertex $x$ it suffices to let $g^\prime(x)$ be $g(x)$ rounded to the nearest value of $\mathcal{L}$, or always rounded down when the nearest larger and nearest smaller values are at the same distance (always rounding up when there is a tie also works).
This clearly maintains the isotonic requirement.
However, for different values of $p$ different proofs of optimality are needed.

For $p=1$ one can always have an optimal real-valued regression where all the values are in the set of values of $f$.
Since we are only concerned with the case where $f$ and $g^\prime$ are $\mathcal{L}$-valued, a $\wellp{1}$ or $\ellp{1}$ regression can be chosen to be $\mathcal{L}$-valued with the same $L_1$ error of an optimal real-valued regression without constraints on the regression values. 
However, it may be that the real-valued regression returned by an $L_1$ algorithm produces a $g$ where not all values are in $\mathcal{L}$.
For example, if $f$ is 1, 0 then $g$ can be of the form $\alpha, \alpha$ for any $\alpha \in [0,1]$.
If $\mathcal{L} = \{0,1\}$ then the values of $g$ must be adjusted.
In general, if the value $x$ of a level set $S$ is not a label value then anything from $\max\{\ell: \ell \leq x, \ell \in \mathcal{L}\}$ to $\min\{\ell: \ell \geq x, \ell \in \mathcal{L} \}$ is an optimal regression value of $S$, so for $\ellp{1}$ and $\wellp{1}$ setting the regression value of $S$ to to be the closest of these, and always rounding down in case of ties, maintains optimality and isotonicity.

For $p=\infty$,
throughout we assume that the basic $L_\infty$ isotonic regression for unweighted data is used:
for any vertex $x$, 
let $\alpha(x) \in \arg\max\{f(u): u \preceq x\}$ and $\beta(x) \in \arg\min\{f(v): x \preceq v\}$.
Then the real-valued regression value $g(x)$ is $(f(\alpha(x))+f(\beta(x)))/2$.
If a regression value $g^\prime(x) =c$ is used then the error at $\alpha(x)$ is $\geq f(\alpha(x))-c$, and the error at $\beta(x)$ is $\geq c-f(\beta(x))$, so minimizing the distance from $c$ to $x$ minimizes the $L_\infty$ error imposed by the value of $g^\prime(x)$.

For $p=2$, for any level set $S$ its optimal regression value is the average of the values in it.
If the regression value is $x$, then for any $\epsilon > 0$, the error of using $x- \epsilon$ as the regression value for $S$ is the same as the error of using $x+\epsilon$, and the error is monotonic in $\epsilon$.
Therefore for an $\mathcal{L}$-valued function the error contributed by $S$ is minimized by setting the regression value to be the closest value of $\mathcal{L}$, or rounding down if there are two closest values.
As before, this also preserves isotonicity.

\subsection{Weak \wellp{\infty} and Strong \ellp{\infty}}  \label{ellpinfty}

Given an $f$-isotonic set $C$, 
for every vertex $v \in V$, $v$'s \textit{C-window} is $[\winlow{v}, \winhigh{v}]$, where
$\winlow{v} = \max\{f(u): u \in C, u \preceq v \}$ (it is $-\infty$ if $v$ has no predecessor in $C$) and $\winhigh{v} = \min\{f(u): u \in C,  u \succeq v\}$ (or $+\infty$ if $v$ has no successor in $C$),
i.e., the window is the range of possible values an isotonic regression can have at $v$, given that the regression values on $C$ are the values of $f$.
If $C$ is a maximum $f$-isotonic set and $v \not\in C$ then $f(v)$ is not in $v$'s window, for if it were then $v$ could be added to $C$ to give a yet larger $f$-isotonic set.
Given $C$, the $C$\textit{-trim} of $f$ at $v$ is the closest value to $f(v)$ in $v$'s $C$-window, i.e., it is $f(v)$ if $f(v)$ is in $v$'s $C$-window, $\winlow{v}$ if $f(v) < \winlow{v}$, and $\winhigh{v}$ if $f(v)> \winhigh{v}$.
The \textit{trim error of $f$ due to} $C$, \trimerr{f,C}, is the maximum, over all $x \in V$, of the distance from $f(x)$ to its $C$-trim.
For a vertex $v$ we slightly abuse notation and use $\trimerr{f,v}$ to mean $\trimerr{f,\{v\}}$.
A single vertex is $f$-isotonic, and
$$\trimerr{f,v} = \max\left\{\{\max\{f(v)\!-\!f(x): v \preceq x, f(v) \geq f(x)\},~\max\{f(x)\!-\!f(v): v \succeq x, f(v) \leq f(x)\}\right\}$$
For an arbitrary $f$-isotonic set $C$, $\trimerr{f,C} = \max\{\trimerr{f,v}: v \in C\}$.

\begin{lemma}   \label{lem:inftytrimbound}
Give a set of real-valued labels, a label function $f$ on a dag $G=(V,E)$, an $f$-isotonic set $C \subseteq V$,
and an $L_\infty$-optimal isotonic regression $g$ of $f$, let $g^\prime$ be $g$ trimmed to $C$.
Then $g^\prime$ is isotonic and $||f-g^\prime ||_\infty = \max\{||f-g||_\infty,\, \trimerr{f,C}\}$.
\end{lemma}
\textit{Proof:}~
The $C$-windows are isotonic in that their lower bounds and upper bounds are isotonic functions.
Trimming an isotonic function to isotonic windows always results in an isotonic function.

To show the upper bound on $||f - g^\prime||_\infty$, for any $x \in V$ suppose $g(x) \geq f(x)$ and let $y \in C$, $x \preceq y$, be such that $\winhigh{x}=f(y)$.
If $g(x) \geq \winhigh{x}$ then $g^\prime(x)=\winhigh{x} \leq \trimerr{f,y}$; if $g(x) \leq \winlow{x}$ then $g^\prime(x)$ is $g(x)$ trimmed up to $\winlow{x} \leq f(x)$ and $|g^\prime(x)-f(x)| < |g(x)-f(x)|$; and if $g(x) \in [\winlow{x},\winhigh{x}]$ then $g^\prime(x) = g(x)$.
Thus in all cases $|g^\prime(x)\!-\!f(x)| \leq \max\{|g(x)\!-\!f(x)|, \trimerr{f,y}\}$, which shows that $||f\!-\!g^\prime ||_\infty \leq \max\{||f-g||_\infty,~ \trimerr{f,C}\}$.
Similar results hold if $g(x) \leq f(x)$, and since both terms in the $\max$ are also lower bounds the equality is proven.
~$\Box$
\medskip

\begin{proposition}  \label{thm:wellpinfty}
Given a set of real-valued labels, a label function $f$ on a dag $G=(V,E)$, and an $f$-isotonic set $C\subseteq V$, an \wellp{\infty} isotonic regression of $f$ can be obtained in $\Theta(m)$ time.
\end{proposition}
\textit{Proof:}~ 
The standard simple isotonic regression $g$ which minimizes $||f-g||_\infty$, namely $g(x) = (\max\{f(y): y \preceq x\} + \min\{f(y): y \succeq x\})/2$, can be computed via topological sort, taking $\Theta(m)$ time.
Using this specific $g$ minimizes the time and the lemma shows it minimizes the error since \trimerr{f,C} is independent of the isotonic function used.
~$\Box$

\medskip

\begin{theorem} \label{thm:ellpinfty}
Given a set of real-valued labels, a label function $f$ on a dag $G=(V,E)$, and a violator graph $\vio{G}=(\vio{V},\vio{E})$,  an \ellp{\infty} isotonic regression of $f$ can be obtained in $\Theta(\mathcal{T}(\vio{n},\vio{m}) \log n)$ time.
\end{theorem}
\textit{Proof:}~
The proof of the theorem shows that to find a $\ellp{\infty}$ isotonic regression $g$ of $f$ we merely need to find an $f$-isotonic set  $C$ of size $s = n-\Delta_\infty(f)$ that minimizes trim-err among all such sets.
We can determine $s$ by just running the algorithm to find an optimal relabeling function of $f$.

To find $C$,
there is an $f$-isotonic set of size $s$ with \terr~ $\leq t$ iff the minimum satisfying flow is $s$ on an adjusted violator graph where for every $v$ where $r(v)>t$, the required flow from $v^\mathrm{in}$ to $v^\mathrm{out}$ is set to 0.
A simple binary search on the trim-err values of the vertices can be used to find the smallest possible $t$.
The regression function used in the lemma can be found in $\Theta(m)$ time, as can $\winlow{\cdot}$ and $\winhigh{\cdot}$, so the time is dominated by the time to find the violator graph and the logarithmic number of iterations to determine a maximal $f$-isotonic set.
~$\Box$
\smallskip

\subsection{Weak \wellp{p} Optimality,~ $p \in \{1,  2\}$}  \label{sec:wellp12}

For \wellp{1}, finding an $L_1$ optimal regression and trimming may not be optimal.
On a linear order, suppose the values are $0^*, 3, 1^*, -1, -2, -3, -4, 2^*$, where $^*$ indicates the unique maximum isotonic set $C$.
The unique $L_1$ optimal isotonic regression has values -1, -1, -1, -1, -1, -1, 2, which, when trimmed, is 0, 0, 1, 1, 1, 1, 2, with $L_1$ error 13.
However, the unique \wellp{1} isotonic regression is 0, 1, 1, 1, 1, 1, 1, 2, with error 12.

A different approach is to trim the results then find the optimal regression of the the trimmed values.
Simple examples shows this does not work for \wellp{\infty}, but it does for \wellp{1}.

\begin{proposition}   \label{thm:L01trimthenopt}
Given a set of real-valued labels, a label function $f$ on a dag $G=(V,E)$, and an $f$-isotonic set $C\subseteq V$, an \wellp{1} isotonic regression can be obtained by trimming $f$ to $C$ and finding an $L_1$ isotonic regression of the trimmed values.
\end{proposition}
\textit{Proof:}~
Any \wellp{1} isotonic regression $g$ must have its values in $[\winlow{v},\winhigh{v}]$ for all $v$, and hence if $f(v) < \winlow{v}$ then the error at $v$ is $(\winlow{v} - f(v)) + (g(v) - \winlow{v})$. 
Similarly if $f(v) > \winhigh{v} $ then the error at $v$ is the distance to the trimmed value,  $\winhigh{v}$, plus the distance from $\winhigh{v}$ to $g(v)$.
Therefore a function minimizing the sum of the distances from the trimmed values to their regression values is also a function minimizing the $L_1$ error of an isotonic regression consistent with $C$, and vice versa.
~$\Box$
\smallskip

Trimming is just a topological sort operation on the original graph, so can be completed in $\Theta(m)$ time.
Since $C$ is given, the total time is determined by the time to find the $L_1$ isotonic regression on the original graph.
The fastest known algorithms depend on the graph~\cite{QFastest}.
The fastest for $d$-dimensional grids, or points in arbitrary position in $d$-dimensional space, is discussed in Section~\ref{sec:multidim}.
The fastest for general dags takes $\Theta(nm+n^2\log n)$ time \cite{L1Iso_AHKW} time
or  $\tilde{O}(\min\{m^{1+o(1)}, m + n^{1.5}\})$ time with high probability~\cite{Flow_BLLSSSW,Flow_CKLPPS}.

For \wellp{2} simple examples show that neither finding an optimal regression and then trimming it (as for \wellp{\infty}) nor trimming and then finding an optimal regression (as for \wellp{1}) always gives an optimal answer.
Instead we borrow a technique appearing in~\cite{Relab_Radeeta09}.
Their technique applies more generally, finding \ellp{2}, though only when regression values are restricted to the labels.
Further, the time of their approach replaces $n$ with $n\ell$, and it is not clear how to utilize modern flow algorithms.

\begin{proposition}    \label{prop:L02}
Given a set of real-valued labels where all labels are integers in $[1,U]$, a label function $f$ on a dag $G=(V,E)$, and an $f$-isotonic set $C\subseteq V$, the \wellp{2} isotonic regression can be found in time linear in the time to do an integer-valued $L_2$ isotonic regression of a weighted function on $G$ when all values and weights are integers in $[0,(n U)^3]$.
\end{proposition}

\noindent\textit{Proof:}\, For sets $S_1, S_2$ of size $\leq n$, with unit weights and where all entries are integers in $[1,U]$, their means (their $L_2$ regressions) are either the same or differ by less than $\alpha = 1/(nU)^2$.
A weight $W$ will be assigned to each vertex in $C$, and the unit weight retained at all other vertices, so that any set containing a vertex $v$ of $C$ will have a mean $x$ less than $\alpha/2$ from $f(v)$.
For arbitrary additional vertices $x-f(v)$ is no larger than $(Wf(v)+nU)/W -f(v) = nU/W$, so choosing $W = 2nU/\alpha$ suffices.

Multiplying all values by $\lceil 1/\alpha \rceil$ guarantees that the integer $L_2$ regression of the scaled values differ by more that $1/\alpha$ if the sets do not intersect $C$, and that any level set containing a member of $C$ will have a regression value which rounds to $f(v)$.
Scaling the values back by dividing by $\lceil \alpha \rceil$ gives the solution to the original problem.
Note that one can then obtain the exact regression values by computing the mean of each level set.
~$\Box$.

Here too the fastest known algorithms depend on the dag, where the fastest algorithm for $L_2$ isotonic regression for general dags currently known is $\Theta(nm \log(m/n))$~\cite{HochbaumQueyranne} for exact regression on arbitrary real-valued functions, or $\min\{\tilde{\Theta}(n^\omega), \Theta(m^{1.5})\}$ with high probability~\cite{LpIso_Yale} for integer-valued functions, where the time includes a factor dependent on $\log U$.
See~\cite{QFastest} for updates on the fastest algorithms for various dags.

\subsection{Weak \wellp{0} Optimality}

Determining \wellp{0} appears to require a different approach, given in Figure~\ref{alg:wellp0}. The time analysis is straightforward.

\begin{figure}
$
\mathsf{\flow{G} =(\flow{V},\flow{E})~~\{the~flow~graph~of~a~violator~dag~\vio{G}=(\vio{V},\vio{E})~for~G\}}\\ 
\mathsf{W=\emptyset~~\{the~initial~set~of~vertices~v~where~their~regression~value~f'(v)~has~been~determined\}}\\
\mathsf{for~d=0,~\ell-1}\\
\hspace*{0.2in}\mathsf{g~=~trim~of~f,~trimmed~using~the~values~of~f'~on~W}\\
\hspace*{0.2in} \mathsf{V'~=~vertices~of~V~where~g~and~f~differ~by~exactly~d~steps}\\
\hspace*{0.2in} \mathsf{G'~=~flow~graph~constructed~from~\flow{G},~where~all~vertices~in~V-V'~are~collapsed}\\
\hspace*{0.2in} \mathsf{C=an~f\!-\!isotonoic~set~of~maximum~size~determined~by~G'}\\
\hspace*{0.2in} \mathsf{for~all~v \in C,~f'(v)=g(v)}\\
\hspace*{0.2in} \mathsf{W=W \cup C}
 $
 \caption{Determining an $f'$ which is an \wellp{0} regression of $f$ on $G$}  \label{alg:wellp0} 
 \hrulefill
 \end{figure}

\begin{theorem}
Given a label function $f$ on a graph G, and a violator dag $\vio{G}=(\vio{V},\vio{E})$, the algorithm in Figure~\ref{alg:wellp0} produces an \wellp{0} regression of $f$ in
$\Theta(\mathcal{T}(\vio{n}, \vio{m})\ell)$ time.
\end{theorem}

Previous authors used $\mathcal{T}=\Theta(n^3)$.
Determining the best flow algorithm to use depends on the relationship of $\vio{n}$ to $\vio{m}$,
see Section~\ref{sec:background}, but in all cases can be a significant improvement on this.
In some cases this is unknown a priori, while in others it is and can be exploited.
See Section~\ref{sec:multidim} for such a case.

\section{Multidimensional Orderings}  \label{sec:multidim}

Given points $x = (x_1,\ldots,x_d)$, $y=(y_1,\ldots,y_d)$, $x \neq y$, in a $d$-dimensional space, $y$ \textit{dominates} $x$ iff $x_i \leq y_i$ for $1 \leq i \leq d$.
Domination is also known as multidimensional ordering, component-wise order, or product ordering.
There is no requirement that the dimensions are the same, merely that each is linearly ordered.
This is an extremely important class of dags since there are a vast number of papers in many different applications which
use such orderings, thus algorithms to handle large datasets with such orderings are quite important.

The edges corresponding to multidimensional ordering are implied, not explicit.
Given a set $V$ of $n$ $d$-dimensional points, simple pairwise comparisons could be used to create an explicit dag, but this would require $\Theta(n^2)$ time and may generate a dag with $\Theta(n^2)$ edges (the time depends on $d$ since comparing points can take time linear in $d$, but this small aspect is usually ignored, or reflected by a statement such as ``where the implied constants depend on $d$\,'').
The violator graph could be constructed the same way, and would have the same problems.
However, more concise graphs are possible by embedding $V$ into a larger dag $\check{G}=(\check{V},\check{E})$, where $V \subset \check{V}$, that preserves the ordering on $V$. I.e., if $x, y \in V$ then $x \prec y$ iff there is a path in $\check{G}$ from $x$ to $y$.
The vertices in $\check{V} \setminus V$ are sometimes known as \textit{Steiner vertices}.
An explicit construction of this appears in~\cite{QMultidim}, where it is shown how to embed into a dag with 
$\Theta(n \log^d n)$ vertices and edges, with the construction taking time proportional to this.
In~\cite{QMultidim} the resulting graph is called a \textit{rendezvous graph} since for any $x \prec y$ there is a unique Steiner vertex $s$, their rendezvous, such that there is an edge from $y$ to $s$ and one from $s$ to $x$
That paper also shows how to construct a compressed rendezvous graph, where one dimension is treated differently, that has $\Theta(n \log^{d-1} n)$ vertices and edges, where again the construction takes time proportional to its size.
Both of these increase the number of vertices compared to the simplest dag representation, but this is more than compensated for by a significant reduction in the worst-case number of edges.

An interesting aspect of this is that the violator graph can be similarly constructed.
Given a label function $f$, create the violator graph by using a $(d+1)$-dimensional ordering on $V$ by adding an extra dimension with value $f(v)$ at each vertex $v$,
where the ordering is reversed at this additional dimension, i.e., $(x,f(x)) \prec (y,f(y))$ iff $y \prec x$ and $f(x) > f(y)$.
Using the compressed rendezvous construction mentioned above constructs a violator graph in $\Theta(n \log^{d} n)$ time, having 
$\Theta(n \log^{d} n)$ vertices and edges.
This is an unusual setting in that both the original graph and the violator graph are nearly linear in size and can be constructed in nearly linear time.

Using this gives:

\begin{theorem}   \label{thm:ellpmutidim}
For any dimension $d\geq 2$, given a real-valued function $f$ on $n$ $d$-dimensional points, an $L_0$ isotonic regression can be found in $\Theta(\mathcal{T}(n \log^{d}n, n \log^{d}n))$ time, and an $\ellp{\infty}$ isotonic regression can be found in time a factor of $\log n$ slower, where the implied constants depend upon $d$.
\end{theorem}
\textit{Proof:}~
As noted above, a violator dag $\vio{G}$ can be constructed in $\Theta(n \log^{d} n)$ time, with $\Theta(n \log^{d} n)$ edges and nodes.
The standard conversion of $\vio{G}$ into a flow graph is used, except that the Steiner vertices are not expanded into a pair and all edges into and out of them have 0 minimal flow.
Using this, a maximum $f$-isotonic set $C$ can be found in the time claimed.
This gives the result for $L_0$.
The result for \ellp{\infty} follows from Theorem~\ref{thm:ellpinfty}.
~$\Box$

\begin{corollary}   \label{cor:wellp1mutidim}
Given a real-valued function $f$ on $n$ $d$-dimensional points, and given an $f$-isotonic set $C$ of maximum size, an \wellp{1} isotonic regression can be found in
\newcounter{bean}
\begin{list}
{\alph{bean})}{\usecounter{bean}}
\item $\Theta(n \log n)$ time if $d=1$,
\item $\Theta(n \log n)$ time $d=2$ and the points form a grid,
\item $\Theta(n \log^2 n)$ time if $d=2$ and the points are in arbitrary position, and
\item $\Theta(n^{1.5} \log^{d+1})$ time if $d \geq 3$, where the implied constants depend on $d$.
\end{list}
\end{corollary}
\textit{Proof:}\,
This follows from Proposition~\ref{thm:L01trimthenopt}, which states that, given $C$, one merely needs to trim and then find an optimal $L_1$ regression.
The fastest known times for the $L_1$ regression appear in~\cite{AhujaOrlin} for a), \cite{QPartition} for b), and~\cite{QMultidim} for c) and d).
~$\Box$.

\begin{corollary}    \label{cor:wellpinftymutidim}
Given a real-valued function $f$ on $n$ $d$-dimensional points, and given an $f$-isotonic set $C$ of maximum size, an \wellp{\infty} isotonic regression can be found in
\begin{list}
{\alph{bean})}{\usecounter{bean}}
\item $\Theta(n)$ time if the points form a grid, and
\item $\Theta(n \log^{d-1} n)$ time if the points are in arbitrary position,
\end{list}
where for both cases the implied constants depend on $d$.
\end{corollary}
\textit{Proof:}~
This follows from Theorem~\ref{lem:inftytrimbound}, which states that one merely needs to find an $L_\infty$ isotonic regression and trim it, coupled with the fact that the basic $L_\infty$ isotonic regression can be determined in time linear in the number of edges.
~$\Box$

\medskip
The following appears in~\cite{QLpviaL0}:
\begin{corollary}    \label{cor:wellp2multidim}
Given a function $f$ on $n$ $d$-dimensional points, where all values are integers in the range $[0,U]$, and given an $f$-isotonic set $C$ of maximum size, an \wellp{2} isotonic regression, where all values must be integers, can be found in:
\begin{list}
{\alph{bean})}{\usecounter{bean}}
\item $\Theta(n)$ time if $d=1$,
\item $\Theta(n \log U)$ time if $d=2$ and the points form a grid,
\item $\Theta(n \log n \log U)$ time if $d =2$ and the points are in arbitrary position,
\item $o(n^{1.5n})$ time if $d\geq 3$, where the implied constants depend on $d$.
\end{list}
$\Box$
\end{corollary}

\section{\ellp{p} Isotonic Regression on Linear Orders}  \label{sec:linearellp}

For linear orders we can efficiently determine \ellp{p} via a left-right scan, rather than constructing a violator graph, but cannot use the PAV (pool adjacent violators) algorithm since it does not hold for $L_0$ error.
E.g., if $\mathcal{L} = \{0, 1, 2\}$ and the data values are 2, 2, 2, 0, 0, 1, 1, then after processing the first 5 values the unique optimal $L_0$ regression is the single level set 2, 2, 2, 2, 2. 
When the 6th is processed the result could be either 2, 2, 2, 2, 2, 2  or 0, 0, 0, 0, 0, 1, and when the last is processed the unique answer is 0, 0, 0, 0, 0, 1, 1.
Thus whichever choice is used at the end of the 6th, then either 5 to 6, or 6 to 7, results in merging adjacent violators but getting 2 level sets, i.e., they are not pooled.

To determine \ellp{p} on linear orders we instead use an approach based on the standard algorithm for maximal nondecreasing subsequences.
At each vertex $i$ we determine a pair $(c,s)$, where $c$ is the size of the largest $f$-isotonic set from 1 to $i$, and $s$ is the smallest error among all such sets.
If the previous best value at $i$ was $(c_1,s_1)$ and now it is determined it can be reached in $(c_2,s)$, 
then the best value at $i$ is given by the maxmin operation, where 
$$
\mathrm{maxmin}\{(c_1,s_1),\, (c_2,s_2)\} = 
    \left\{   \begin{array}{ll}
    (c_1,s_1) & \mathrm{if~} c_1 > c_2\\
    (c_2,s_2) & \mathrm{if~} c_1 < c_2\\
    (c_1,\min\{s_1,s_2\}) & \mathrm{if~} c_1 = c_2
    \end{array}   \right.
$$

\noindent We will prove

\begin{theorem}    \label{thm:linearreal}
Given a real-valued label function $f$ on a linear ordering of length $n$, where there are $\ell$ different initial label values and regression values can be arbitrary real numbers,
an \ellp{p} isotonic regression can be found in 
$\Theta(n \ell)$ time for $p \in \{1, 2\}$ and $\Theta(n \log \ell)$ time for $p = \infty$.
Further, the same times can be achieved if all initial and regression values must come from a real-valued label set $\mathcal{L}$ of size $\ell$.
\end{theorem}
\noindent  and

\begin{theorem} \label{thm:linearellp0}
Given a label function $f$ on a linear ordering of length $n$, where the initial and regression values are in a label set $\mathcal{L}$ of size $\ell$, 
a strong $L_0$ isotonic regression can be found in $\Theta(n \ell^2)$ time.
\end{theorem}

\noindent
Since the ordering is linear we assume the vertices are $1, \ldots, n$.

\subsection{$p=1$} \label{sec:linear1}

For \ellp{1} there is a quite simple algorithm based on the fact that the optimal value of a level set can be taken to be a median value of the set, which implies that it can be chosen to be one of the values of $f$.
We can then assume that we use a label set $\mathcal{L} = \{\lambda_1, \ldots \lambda_\ell\}$ which is these values.
For a $(c,s)$-valued array $\mathsf{A(1\!:\!n, 1\!:\!\ell)}$, let $\mathsf{A(i,j)}$ be the optimal \ellp{1} regression of $f$ on $[1,i]$ among those with value $\lambda_j$ at $i$.
Then
$$
\mathsf{A(i,j) = maxmin\{A(i\!-\!1,j): j\leq i\} }+
    \left\{   \begin{array}{ll}
     (0,|f(i)-\lambda_j|) & \mathrm{if~} f(i)\neq \lambda_i\\
     (1,0)                      &  \mathrm{if~} f(i) = \lambda_j 
    \end{array}   \right.
$$

This can easily be computed in $\Theta(n\ell)$ time. An \ellp{1} isotonic regression is one with an optimal maxmin value in $A(n,\cdot)$ and it is straightforward to determine the regression.

\begin{figure}

\centerline{\scalebox{0.85}{\includegraphics{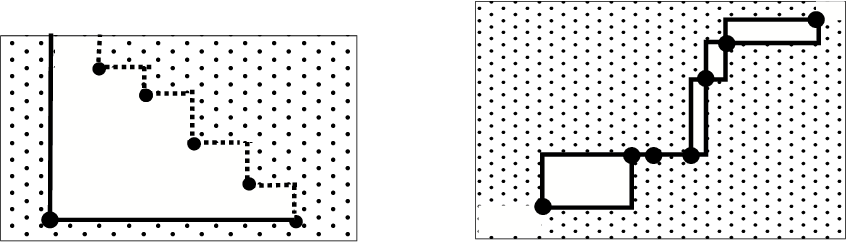}}}
\medskip
\begin{quote}{\small There are no data points in the blank areas (including the lower left and upper right on the right figure) or on the boldface lines (other than vertices of C or possible successors).
For both figures there might be data points in the background, or for the left figure perhaps on the dotted lines (points there aren't successors).}
\end{quote}
\vspace{-0.1in}
\caption{~~~Left: A point (lower left) and all of its possible successors. ~~~~~Right: C for some data set.}
\label{fig:successorgraph}

\hrulefill
\end{figure}

\subsection{$p=2$} \label{sec:linear2}

For \ellp{p}, $p>1$, if the isotonic regression is restricted to values in $\mathcal{L}$ the same approach as for $p=1$ can be used, but when regression values can arbitrary real numbers a more complex approach is needed.
Suppose an \ellp{p} isotonic regression $f'$ is equal to $f$ on an $f$-isotonic set $C = \{i_0 < \ldots < i_k\}$ of maximum size, for some $k \geq 0$.
For $0 \leq j < k$ consider the closed rectangle with lower left coordinate $(i_j, f(i_j))$ and upper right coordinate $(i_{j+1}, f(i_{j+1}))$.
Except at these corners there cannot be any point $(x,f(x))$ in the rectangle for if there were then $C$ would not have maximum size since the point could be added to it.
Thus if we have an optimal $f$-isotonic regression from 0 to $i$, and at $i$ the regression value is $f(i)$, then for the next index $j$, in increasing order, where the regression value is the same as $f(j)$ we need only consider those $j$ where $f(j) > f(i)$, and $f(j) < f(k)$ for all $i+1 \leq k < j$.
To simplify notation, at a vertex $i$ the \textit{potential successors} is the unique sequence of vertices $i < m_1 < \ldots <m_k$ of maximum length where $f(m_1) > f(i)$ and $f(m_1) > f(m_2) > \ldots > f(m_k)$.
See Figure~\ref{fig:successorgraph}.
When we refer to the $L_p$ error of a level set on $[i,j]$ we actually mean the sum of the $p^\mathrm{th}$ power of the pointwise regression errors  using a constant function which minimizes the error among trimmed regression values permitted (if no trimming is used then it would be the $L_p$ mean).
This observation immediately gives the algorithm in Figure~\ref{alg:linearreal}.

\begin{figure} 
$
\mathsf{array~T(0:n\!+\!1)~of~(c,s)~pairs~~\{also~keep~track~of~predecessor~giving~optimal~value\}}\\
\mathsf{f(0)~=~-\infty}\\
\mathsf{f(n+1)~=~\infty}\\
\mathsf{for~j=0,n+1}\\
\hspace*{0.2in}\mathsf{T(j) = (0,0)}\\
\mathsf{for~i=0,n}\\
\hspace*{0.2in}\mathsf{successorvalue=\infty,~~initial~level~set~value~f(i)}\\
\hspace*{0.2in}\mathsf{for~j=i+1, n}\\
\hspace*{0.4in}      \mathsf{if~((f(j)< f(i))~\vee~(f(j) \geq successorvalue))~then~~\{j~is~not~a~sucessor\}}\\
\hspace*{0.6in}         \mathsf{add~level~set~for~j}\\
\hspace*{0.6in}         \mathsf{merge~level~sets~on~[i,j]~trimmed~to~[f(i), successorvalue]}\\
\hspace*{0.4in}      \mathsf{else~~\{j~is~a~sucessor\}}\\
\hspace*{0.6in}         \mathsf{T(j) = maxmin\{T(j), T(i)+(1,error~of~level~sets~on~[i,j-1]~trimmed~to~[f(i),f(j)])}\\
\hspace*{0.6in}         \mathsf{if~(f(i)=f(j))~then~exit~for\!\!-\!\!loop}\\
\hspace*{0.6in}         \mathsf{successorvalue=f(j)}\\
 $
\caption{Determining an \ellp{p} Isotonic Regression on a Linear Order, $p<\infty$}  \label{alg:linearreal}
\hrulefill
\end{figure}

To use this algorithm to prove Theorem~\ref{thm:linearreal} for $p=2$ we need to be able to efficiently merge the level sets and determine the optimal error subject to the trimming requirement.
As in the standard prefix approach to isotonic regression, if a level set has a value $b$ and its predecessor has value $a$, where $a \geq b$, they are combined and the value of the result is calculated (along with ancillary numbers such as the sum of the $f$ values).
Before any calculations of $L_2$ errors we first compute $S_1(j) = \sum_{k=1}^j f(k)$ and
$S_2(j) = \sum_{k=1}^j f(k)^2$ for $1  \leq j \leq n$, and let $S_1(0)=S_2(0)=0$.
In the prefix approach for $L_2$ isotonic regression each level set's regression value is just the average of its $f$ values.
If this is $c$, and the level set is trimmed to $[a,b]$, then $c$ is used as the level set's regression value if $c \in [a,b]$, while otherwise it is $a$ if $c <a$ and $b$ if $c>b$.
For a level set on $[i,j]$, the square of the $L_2$ error of using $d$ as the regression value is \vspace*{-0.05in}
$$S_2(j)-S_2(i\!-\!1) -2 d (S_1(j)-S_1(i\!-\!1)) + d^2 (j-i+1)  \vspace*{-0.05in}$$
so calculating a trimmed level set's error, even when the mean is not used as the regression value, takes constant time.

This shows that for any $i$, the total time to find the maxmin values for all successors is linear in the time to find the last successor, or to $n$ if there is no later $j$ value where $f(i)=f(j)$.
Thus for any fixed $\lambda$, the worst-case time to determine successor values for all $i$ where
$f(i)=\lambda$ is $\Theta(n)$ and the total time is $\Theta(n \ell)$.

\subsection{$p=\infty$}   \label{sec:linearinfty}

There is a very simple $\Theta(n \log \ell)$ time algorithm for $p=\infty$.
First determine the trim error of every point, taking $\Theta(n)$ time.
Then do the standard algorithm for finding a longest nondecreasing sequence, inserting points into a balanced tree, where each node stores the length of the longest nondecreasing decreasing sequence reaching that point, and the minimum maximum trim error of such a sequence.
When a new point is inserted it adds 1 to the length of its predecessor, and computes the max of its trim error and the predecessor's.
A successor point is eliminated iff it has length less than the new point, or the same length and equal or greater trim error.
\medskip

This completes the proof of Theorem~\ref{thm:linearreal}. ~$\Box$

\subsection{Strong $L_0$}   \label{sec:linear0}

An algorithm for finding a strong $L_0$ isotonic regression can be based on that for $L_{0,1}$.
Given an isotonic function $g$ on $[1,m]$ let $v = (v_0,v_1,\ldots,v_{\ell-1})$ be a vector where $v_i$ is the number of entries of $g$ which are $i$ labels away from the values of $f$ on $[1,m]$.
Recall that a strong $L_0$ isotonic regression of $f$ on $[1,n]$ maximizes $v_0$, and among the functions that maximize $v_0$ maximizes $v_1$, etc.
Let $\mathsf{A(i,j)}$ be a vector-valued array, $1\leq i \leq n$, $1 \leq j \leq n$, where $\mathsf{A(i,j)}$ is a vector corresponding to a strongest isotonic regression of $f$ on $[1,i]$, among those that are $\lambda_j$ at $i$.
For vectors $u, v$ of length $n$ let $\mathrm{vmax}(u,v) = u$ if there is an $i$ such that $u(k) = v(k)$ for $k<i$ and $u(i) > v(i)$, and $v$ otherwise.
Then
$$
\mathsf{A(i,j) = vmax\{A(i\!-\!1,j): j\leq i\} +
     (v_0, \ldots, v_{|f(i)-\lambda_j|}\!+\!1, \ldots, v_{\ell-1})}
$$
A strong $L_0$ isotonic regression optimizes $\mathsf{A(n,\cdot)}$ and can be computed in $\Theta(n \ell^2)$ time.
This completes the proof of Theorem~\ref{thm:linearellp0}. ~$\Box$

\section{$L_p$ Isotonic Regression with Hamming Distance Penalty}   \label{sec:penalty}

In some estimation problems the objective is similar to \ellp{p}, namely, given a function $f$ and constant $\alpha > 0$, find an isotonic function $g$ that minimizes $||f\!-\!g||_p + \alpha ||f\!-\!g||_0$, i.e, an $L_p$ isotonic regression with an $L_0$, or Hamming distance, penalty function.
It is a form of regularized regression.
Note that for $\alpha$ sufficiently large (relative to $f$) this is the same as \ellp{p}.
For $p \in \{1, 2\}$ algorithms are given for the case where the dag is linear, while for $p = \infty$ an algorithm is given for arbitrary dags.

\subsection{$p \in \{1,2\}$}    \label{sec:penalty12}

Here we use the ideas in Section~\ref{sec:linearellp}, but for each possible regression value we no longer keep track of the number  of vertices where the regression had the same value as the original function, but rather just the minimum total cost of reaching that value.

\begin{proposition}   \label{thm:penalty12}
Given a real-valued function $f$ on a linear ordering of length $n$ and a constant $\alpha > 0$, a real-valued isotonic function $g$ which minimizes $||f\!-\!g||_p + \alpha ||f\!-\!g||_0$ among all isotonic functions can be found in $\Theta(n^2)$ time for $p \in \{1, 2\}$.
\end{proposition}

For this problem one can view the set of labels $\mathcal{L}$ to be the values of $f$, and thus there may be $n$ labels.
For $p=1$ the algorithm for \ellp{1} in Theorem~\ref{thm:linearreal} can be applied here by merely changing the recurrence to
$$\mathsf{A(i,j) = \min\{A(i\!-\!1,j): j\leq i\} +\alpha |f(i)-\lambda_j| \cdot (f(i)\neq \lambda_j)}
$$
Thus the time for an $L_1$ penalty can be expressed as $\Theta(n\ell)$, where $\ell$ is the number of different initial values.

For $p=2$ the algorithm is the same as that in Theorem~\ref{thm:linearreal} for \ellp{2}, except that here at each potential successor 
one merely keeps track of the minimum total cost of getting there.

\subsection{$p = \infty$}  \label{sec:penaltyinfty}

We give a result which is more general than linear orders:

\begin{theorem}   \label{thm:penaltyinfty}
Given a real-valued function $f$ on a dag $G(V,E)$ of $n$ vertices, let $k=n-\Delta_0(f)$.
Then given a constant $\alpha > 0$, a real-valued isotonic function $g$ which minimizes $||f\!-\!g||_\infty + \alpha ||f\!-\!g||_0$ among all isotonic functions can be found in
\begin{list}
{\alph{bean})}{\usecounter{bean}}
\item $\Theta(nk \log(n/k))$ time if $G$ is a linear order
\item $\Theta(k \log(n/k) \mathcal{T}(\vio{n},\vio{m}))$ time for an arbitrary dag if a violator graph $\vio{G}$ is given.
\end{list}
\end{theorem}

\noindent
We assume $k >0$.
One can first check if the data is isotonic, taking $\Theta(m)$ time, and if it is then $g = f$, the total error is 0, and $k=0$.
If the data isn't isotonic then $k > 0$.

If $\alpha$ is sufficiently large then an isotonic function $g$ is in \ellp{\infty}$(f)$ iff it minimizes $||f\!-\!g||_\infty + \alpha ||f\!-\!g||_0$. 
Theorem~\ref{thm:ellpinfty} shows an \ellp{\infty} regression can be found by first determining $k = n - \Delta_0(f)$ and then finding the smallest \terr~ $t$ such that there is an $f$-isotonic set $C$ of size $k$ among the vertices $v$ with 
$\trimerr{v} \leq t$.
Then $f$ trimmed to $C$ is a \ellp{\infty} isotonic regression.
Further, $t$ can be found by a binary search among the \terr~ values, where at each step one uses a flow algorithm to find a largest $f$-isotonic set among the vertices with \terr~ $\leq t$.

The same approach can be used to find an isotonic function $g$ minimizing $||f\!-\!g||_\infty + \alpha ||f\!-\!g||_0$.
For an isotonic function $g$ let $C(g)$ be the set of vertices where $g$ and $f$ have the same value.
Then $||f\!-\!g||_0 = i$ for some $1 \leq i \leq n$ and $g$ must have the property that \trimerr{C(g))} is the minimum \terr~ over all isotonic functions $h$ such that $||f\!-\!h||_0 = i$ ($g$ may not be unique).

This gives a simple way to minimize $||f\!-\!g||_\infty + \alpha ||f\!-\!g||_0$: for each $1 \leq i \leq n$ let $g(i)$ be an isotonic function with minimal \terr~ among those functions $g^\prime$ where $|C(g^\prime)| = n-||f-g^\prime||_0$.
Then a function minimizing $\min\{||f-g(i)||_\infty + \alpha ||f-g(i)||_0: 1 \leq i \leq n\}$ minimizes $||f-h||_\infty + \alpha ||f-h||_0$ among all isotonic functions $h$.

The primary difference between Theorem \ref{thm:penaltyinfty} a) and b) is the technique used to determine the minimum \terr~ needed to achieve $|C(g(i))|=i$.
For both we first determine $\Delta_0(f)$ via the relevant regression algorithm.
We also sort points by their \terr~ value.
\bigskip

\noindent \textit{Proof of a):}~
For the linear order find an optimal $L_\infty$ regression $h$.
Then start with an empty linear order and add points one at a time, keeping them ordered by their order in $V$. After each insertion we determine the size of the longest increasing sequence, and when this value changes from $i-1$ to $i$ we record the \terr~ where this occurred.
After all points have been added we determine the optimal $i$ value, create a linear order of all points with \terr~ $\leq$ the \terr~ needed to obtain this, find a maximum length nondecreasing subsequence $C$ and trim $h$ to this.
Note that we could have started by immediately inserting all points $v$ where $\trimerr{v} \leq ||f-e||_\infty$.

Sorting the vertices by their \terr~ and inserting them one at a time and determining the longest increasing subsequence is called the \textit{dynamic longest increasing subsequence} problem and has been studied by several authors~\cite{LIS_SublinApproxMultidim,LIS_SublinApproxPosets,LIS_SublinApproxDyn1D_21v2} in the more difficult setting where there can be deletions as well as insertions.
When there are only insertions a $\Theta(nk \log(n/k))$ algorithm appears in \cite{LIS_ExactDyn1D_13}.
A randomized algorithm in~\cite{LIS_SublinApproxDyn1D_21v2} for the fully dynamic problem takes $\tilde{\Theta}(n^{5/3})$ time but has a very small chance of producing an incorrect answer (note that they call this exact though it is not always correct).
Here we only do insertions and don't need to produce an LIS every step, just determine its length, so it is likely that an exact algorithm taking $o(n^2)$ time exists.
This would immediately improve the time for a).
\bigskip

\noindent \textit{Proof of b):~}
Insert the points into the violator graph where
for any points not currently inserted we set their minimum flow requirement to 0.
We determine minimum flow in this adjusted graph, which plays the same role as finding the maximum nondecreasing sequence for the linear order.

The difference is that we don't insert all of the points in sequential order by their \terr~ value.
That would result in an algorithm taking $\Theta(n \mathcal{T}(\vio{n},\vio{m}))$ time.
We reduce this slightly, replacing $n$ with $k \log(n/k)$, by using the fact that we are searching for $k$ values and these values are monotonic in the values of \terr.
This is equivalent to searching for a set $S$ of $k$ unknown values in an ordered set of $n$ values, where at each probe one can determine if there are any smaller elements of $S$.
To find them all there are essentially $k$ binary searches in contiguous regions with unknown boundaries.
Since the log function is concave this is maximized when the regions are all the same size, so the worst-case number of probes is $\Theta(k \log(n/k))$.

This finishes the proof of b), which finishes the proof of the theorem. ~$\Box$

\section{Final Remarks}  \label{sec:final}

Datasets are quickly becoming vastly larger and more complex, often with order-constrained relationships between independent and dependent variables, but the relationships are difficult to evaluate due to noise.
Because of this, nonparametric regression, particularly isotonic regression, is becoming increasingly important.
Various criteria may be used for deciding how to enforce isotonic (monotonic) assumptions, and for linearly ordered labels one criterion is to minimize the number of labels that must be changed.
When optimizing this $L_0$ distance (aka Hamming distance), in general there will be many $L_0$ isotonic regressions and choosing among them can be useful but difficult.
We choose ones which optimize secondary criteria based on other $L_p$ metrics, an approach also used in~\cite{Relab_Radeeta09}.
We gave algorithms for computing them, and many authors have already shown the utility of isotonic regression for $L_0$ and other metrics~\cite{MonoDataSci_DeBaets19,Brabantetal,LearnRelab_Cano18,LIS_SublinApproxMultidim,Relab_Feelders06,LIS_SublinApproxPosets,LIS_SublinApproxDyn1D_21v2,Relab_PijlsPot14,Relab_Radeeta09,Relab_Radetal12,RULEM17}.

The algorithms utilize maximum flow algorithms, and for  \wellp{1} and \wellp{2} also utilize algorithms for $L_1$ and $L_2$ isotonic regression.
Since there continues to be advances in algorithms for these problems, and we just use them via subroutine calls, we have given algorithms stated in general terms, using $\mathcal{T}(n,m)$ to represent the fastest maximum flow algorithm relevant to the problem, and general references to the currently fastest $L_1$ and $L_2$ algorithms.
The wikipedia page~\cite{Flow_wiki} is usually up-to-date on the fastest flow algorithms, and~\cite{QFastest} is usually up-to-date on the fastest $L_p$ isotonic algorithms for various dags and $p \in \{0, 1, 2, \infty\}$.

Finally, we introduced strong $L_0$ isotonic regression (Section~\ref{sec:secondary}), aka \ellp{0}, which appears to be a natural choice among $L_0$ isotonic regressions with arbitrary labels, either numeric or nonnumeric.
We gave an efficient algorithm for determining it on linear orders, but not for general dags nor even multidimensional ones.
It is very similar to the \textit{strict $L_\infty$ isotonic regression} in~\cite{QStrict}.
There a $\Theta(\min\{nm, n^\omega\} + n^2 \log n)$ time algorithm was given for determining it, and one taking $\Theta(nm)$ expected time appears in~\cite{LpIso_Yale}.
Given an isotonic regression $g$ of $f$, let $(e_1, e_2, \ldots, e_n)$ be the values $\{|g(v)-f(v)| : v \in V\}$ sorted in decreasing order.
A strict $L_\infty$ isotonic regression is one which is first in the lexical ordering of all sequences corresponding to isotonic regressions.
It may not be unique.
Recall that strong $L_0$ isotonic regression corresponds to a lexically first vector when the $e_i$ are sorted in increasing order.
Strong $L_0$ maximizes the number of small errors while strict $L_\infty$ minimizes the number of large ones.

\end{document}